# Semi-automatic identification of counterfeit offers in online shopping platforms

*Christian Wartner, Patrick Arnold and Erhard Rahm*

## Abstract

Product counterfeiting is a serious problem causing the industry estimated losses of billions of dollars every year. With the increasing spread of e-commerce, the number of counterfeit products sold online increased substantially. We propose the adoption of a semi-automatic workflow to identify likely counterfeit offers in online platforms and to present these offers to a domain expert for manual verification. The workflow includes steps to generate search queries for relevant product offers, to match and cluster similar product offers, and to assess the counterfeit suspiciousness based on different criteria. The goal is to support the periodic identification of many counterfeit offers with a limited amount of manual effort. We explain how the proposed approach can be realized. We also present a preliminary evaluation of its most important steps on a case study using the eBay platform.

# Introduction

From a legal point of view, counterfeiting or product piracy refers to infringements against intellectual property rights, such as copyrights, trademarks and design rights (Organization of Economic Development 2007). Goods that violate these rights or which are purposely produced to defraud potential customers are called counterfeits, copies, imitations, knock-offs or fakes.

The volume of product counterfeiting has substantially increased in the recent past. According to (European Commission 2013), the number of registered counterfeit cases at the European Customs almost increased by a factor of 18 since 2001, especially for small parcels in express and postal traffic that most likely result from internet sales. The Int. Chamber of Commerce ([www.icc-ccs.org/icc/cib](www.icc-ccs.org/icc/cib)) estimates that counterfeiting accounts for about 5 - 7% of world trade, i.e. about $600 billion a year. Almost all kinds of products are subject to counterfeiting, ranging from electronic devices and apparel to food and drugs. Fake products are offered and sold in numerous online shops and auction sites as well as on B2B marketplaces for wholesale trading. Counterfeits not only cause an enormous economic loss, but can also damage the reputation of a company. Buyers of faked products not only receive a low-quality product in many cases, but may even be exposed to serious safety and health risks, e.g., in the case of faked medication (Rahm 2014).

In this study, we focus on product imitations sold on the web without considering infringements of digital content such as software, music or videos. Taking actions against counterfeiting on the web is challenging due to the huge number of involved traders, websites and products. As the people behind product offers on the web are largely anonymous, counterfeiters can remain nameless and faceless. When counterfeit offers are detected and banned from a site, it takes little effort for an infringer to reappear on another site or under a new name (Roth 2011). Traditional efforts to fight counterfeiting include the use of technical means for product authentication such as holograms or RFID tags (Staake,

Thiesse and Fleisch 2005) (Jordan and Kutter 2012). While successful, these methods are not universally applicable or too expensive to control, e.g., for consumer products such as cosmetics and drugs. For detecting counterfeits in web shops, these techniques become even less effective, because an online customer cannot verify the hologram or RFID tag of a product. Furthermore, counterfeiters often use the bait-and-switch strategy where the original image and description of a product are displayed on the website, yet an imitation is delivered (Mavlanova and Benbunan-Fich 2011).

Manually monitoring a large number of websites and searching for faked products is much too laborious and expensive. Hence, we see a strong need for automated methods to identify likely counterfeits and we will propose such an approach. While the counterfeit candidates still have to be manually verified, the proposed approach is expected to significantly reduce the time and cost for detecting counterfeits. There are different operative and legal measures against detected counterfeiters: removing the offers infringing someone's copyright or trademark, closing the account of the counterfeiter from the platform, or suing the counterfeiter. While legal actions are complex and time-consuming, the removal of counterfeits and counterfeiters are relatively easy as first steps and suitable for mass application.

The need to automatically monitor web shops and other e-commerce platforms for counterfeits has been observed by others (e.g. (Pinsdorf and Ebinger 2005) (MarkMonitor 2012)), and there are also some companies specializing on this task. E-commerce sites like eBay also try to identify counterfeits on their platform[1]. However, we are not aware of any publications describing in some detail how to address the challenging task of automatically identifying likely counterfeits.

---

[1] http://pages.ebay.com/againstcounterfeits/

There are various parties who could benefit from a semi-automated monitoring and detection of counterfeits. Their goals differ in the number and type of products and in whether they are interested in only one or several online sales platforms:

1. Manufactures (owners of the trademark / copyright) are only interested in detecting imitations of their own products. They likely have a specific list of their products and try to find counterfeits either within a specific online sales platform or across several such platforms.

2. Public authorities (e.g., customs, police) can be interested in all imitations of one or several kinds of product(s), either within a specific platform or across several platforms.

3. Owners of an online sales platform are interested in detecting all fake products of one or several product type(s) or manufacturer(s) distributed only on their own site.

4. Consumers want to know if the offer they are interested in is genuine or not.

In this study, we focus on describing and analyzing a semi-automatic approach to identify likely counterfeits on a single platform Part 2 explains the proposed workflow and its main techniques to find and cluster relevant offers and to score their trustworthiness. In Chapter 3 and Chapter 4 we present the setup for a case study on the eBay auction platform to show the applicability of the presented workflow and the results of our experiments. Finally we discuss the results and possible future work and research directions.

## A semi-automatic Approach to Detect Counterfeit Offers

### Workflow for analyzing and scoring product offers

We propose a semi-automatic workflow to determine offers for specific products on e-commerce platforms to analyze and score their counterfeit suspiciousness. *Figure 1* shows the main steps in the proposed workflow that will be discussed below. We assume that a user specifies in an input file or selects interactively the products of interest. Depending on her requirements she may define specific products or broad product groups. She could also specify different web data sources (auction sites, web shops, etc.) to be examined but we will focus on a single site in this study.

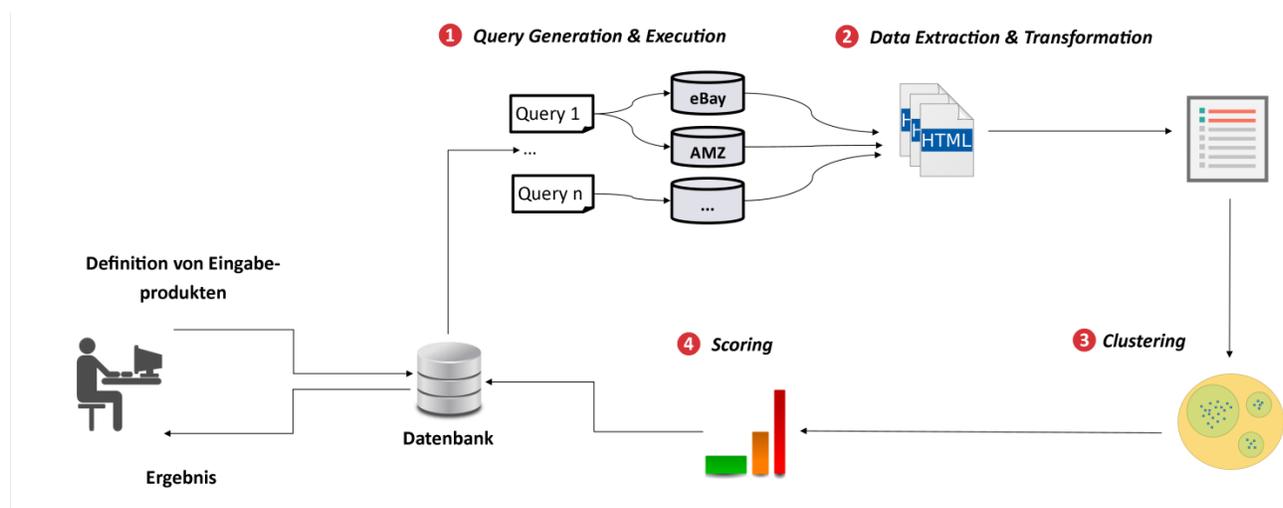

*Figure 1 Workflow for detecting counterfeit offers*

The workflow steps are executed once or periodically and results can be stored in a database from which reports about potential counterfeits can be generated depending on the application needs. The main steps are the following:

1. Querying product offers: The first step is to query or search all product offers that correspond to the specified products of interest. This step entails the automatic generation of a potentially large number of suitable search queries tailored to the search capabilities of the sales platform.

2. Extraction and transformation of product offers: The search results need to be processed to extract the individual product offers with their relevant attributes (e.g., product id, title, retailer, price, date, or customer rating depending on the data source). The product offers are pre-processed and cleaned for the subsequent steps.

3. Clustering of product offers: The retrieved product offers are clustered so that each cluster contains all offers of a specific product. This enables a comparison between different offers of the same product and can thus help to identify suspicious variations.

4. Counterfeit Analysis and Scoring: In the last step, we use different criteria to derive for each product offer a counterfeit score to indicate its likelihood of referring to a fake product. Suspicious offers are shown to the user for manual verification.

In the following we discuss these steps in more detail.

## Query generation

The basis for finding offers for counterfeit products is the search and extraction of product offers from online shops. Searching for counterfeits is based on the specification of products of interest. These may be specific products or all products of a specific type and/or from a specific manufacturer. Manually finding all relevant offers is generally infeasible so that we use the automatic generation of suitable search queries based on the supported query capabilities of the considered online sales platform. Typically, the search interface supports different query predicates such as keyword search in the product title and description, as well as searching for specific attributes such as manufacturer or product type.

The challenge is to utilize these capabilities such that all relevant offers are found with a minimum of queries. This is complicated by the typically highly heterogeneous descriptions and frequently missing information in product offers. Furthermore, there is an inherent trade-off between recall and precision. So specific product queries using manufacturer-specific product codes (e.g., "Gucci - 3509/S") or even global product identifiers such as the EAN or UPC [2] usually return precise results, but can lead to poor recall as this information is often not present in offers at platforms such as auction sites. On the other hand, more general queries for the product type or manufacturer (e.g., "Gucci" in the category sunglasses) have a better chance of returning all relevant offers, but will also return a large number of irrelevant offers that have to be filtered out afterwards. Hence, the precision for such queries is often rather low.

Putting these criteria together requires a system for a customized search for product offers in online sales platforms. According to the requirements of the user, the system must formulate search queries to retrieve the specified products by using the search interface or web API. At best, no relevant products are missing in the result set and no irrelevant products are included. Also, the result should not contain duplicates, i.e., the same offers occurring twice or more. Formulating such queries is called *query generation*. The implementation of this task is challenging but can build on recent research results, e.g. in the area of mashup applications that query information at run time, (Endrullis, Thor and Rahm 2009) (Barbosa and Freire 2004). Some of these approaches have already been applied for finding product offers, e.g. to monitor and compare their prices (Wartner and Kitschke 2011) (Endrullis, Thor and Rahm 2012). For a list of specific products, these approaches can either generate a single query per product, e.g. using information from the product title specified in the input, or it can be tried to find several products in a single query, e.g. if they share the same manufacturer or product type.

## Data extraction and transformation

After queries are generated and executed, the results are stored and the relevant data has to be extracted from the resulting web pages of the shop. Data extraction is easy if the website provides an API to programmatically submit queries and retrieve query results including relevant attributes as title, product id, price, seller, user rating etc.. Otherwise web scrapers must be employed to extract the relevant information from the HTML code of the query result pages. This process can be error-prone and implies a higher effort of creating and maintaining web wrappers.

---

[2] EAN – European Article Number, UPC – Universal Product Code

The extracted product offers generally need further data transformation and cleaning to ensure sufficient quality for the further workflow steps of clustering and analysis. There are different techniques to remove irrelevant or invalid product offers, such as privately owned products in auction sites or offers that lack basic information like the price or title. Some product groups may require specific transformation steps to better support the comparison of their offers. For example, perfume items are typically sold in different sizes ranging from sample sizes of 5 ml up to large packages of 250 ml or more. It is thus important to extract the respective size and perhaps also to compute a normalized price, e.g., the price per ml, for easier price comparison. Other product type specific preprocessing steps are the unification of clothing sizes and mapping synonyms to a single representation (e.g. "bag", "evening bag", "leather bag", "hand bag" → "purse" in the accessories domain).

## Clustering of product offers

The next step in the workflow is the assignment of the extracted offers to specific input products or the clustering of equivalent product offers. This is necessary since we want to use a counterfeit detection approach that considers differences between offers for the same product in addition to individual offer attributes. Clustering or matching of offers is challenging due to their high heterogeneity and missing information. For example, product titles in offers for the same product may differ considerably. For a product officially named "Gucci Sun Dream MD-120b", found offer titles may include "SunDream MD120b" or "Blue Gucci Sun Dream purse, model 120b". Some offer titles may only say "Nice Sun Dream purse from Gucci" thereby lacking significant product details.

Determining offers referring to the same product is a special case of object matching (or entity resolution) which aims at finding equivalent data objects in a dataset referring to the same real-world entity. This problem has been intensively studied already and there many available match approaches typically utilizing a combination of similarity scores for different attributes (Elmagarmid, Ipeirotes and Verykios 2007). For example, one can evaluate the lexicographic similarity of the object names or other attributes. This way, an object matcher can easily recognize that two objects "Blue Gucci, MD120b" and "Gucci MD-120b (blue)" are highly similar so that they likely represent the same object. The use of dictionaries and thesauri helps to discover synonyms and to deal with homonyms. As pointed out in in (Köpcke, et al. 2012), matching product offers is especially challenging and needs tailored approaches by considering information about the manufacturer and product type.

For our purposes, we not only have to decide whether two offers refer to the same product but we want to group all matching offers for the same product in one cluster. Only then it is possible to compare all relevant offers with each other, e.g., with respect to their price. To cluster similar product offers, we propose the use of a hierarchical bottom up clustering. The main advantage of this approach is that the number of elements in each cluster does not have to be known upfront and the clusters do not overlap, which is the case with other state-of-the-art clustering approaches such as k-Means.

The proposed clustering algorithm is illustrated in *Listing 1*. Initially, each element (product offer) represents a separate cluster. We call the algorithm with the initial list of clusters and a minimal similarity threshold to be met by all pairs of elements in a cluster. The algorithm iteratively determines for each current cluster *l* the cluster *j* with the highest similarity. For this purpose, we define the similarity between two clusters as the smallest similarity between any two elements from the different clusters (function CalculateSimilarity). If the highest similarity between clusters *l* and *j* is above the minimal similarity threshold the two clusters are merged since their offers likely refer to the same product. This process is continued as long as there are further pairs of clusters that can be merged.

| 01 | **INPUT: Clusterlist L, minimal pairwise similarity** |
| --- | --- |
| 02 | **FOR EACH Cluster I in L** |
| 03 | **Determine cluster j in L with largest similarity s** |
| 04 | **IF s > minimal pairwise similarity** |
| 05 | **THEN** *mergeClusters( i, j)* |
| 06 | **ELSE quit** |
| 07 | **END IF** |
| 08 | **END FOR** |
| 09 | **FUNCTION mergeClusters( Cluster i, Cluster j)** |
| 10 | **Cluster k = cluster i ∪ cluster j** |
| 11 | **Insert k in L** |
| 12 | **Remove i, j from L** |
| 13 | **END** |
| 14 | **FUNCTION calculateSimilarity( Cluster i, Cluster j)** |
| 15 | **minConfidence = 1** |
| 16 | **FOR EACH x in i and y in j** |
| 17 | **Caclulate confidence c of (x, y) using a match tool** |
| 18 | **IF c < minConfidence** |
| 19 | **THEN minConfidence = c** |
| 20 | **END IF** |
| 21 | **END FOR** |
| 22 | **RETURN minConfidence** |
|    | **END** |

*Listing 1 Pseudo code for hierarchical clustering*

The table below shows a simple scenario with two clusters X, Y, each containing two elements $x_1$, $x_2$ resp. $y_1$, $y_2$, that are compared with each other. Thus, we have to compare the similarity between each pair of cluster elements. The minimum value is 0.93 (similarity between $x_2$ and $y_2$), so we say that the similarity between these two clusters is 0.93. If this similarity exceeds the minimal similarity threshold and if no other cluster pair has a score above 0.93, we would combine the two clusters X and Y.

|   |   | X |   |
|---|---|---|---|
|   |   | $x_1$ | $x_2$ |
| Y | $y_1$ | 1.0 | 0.98 |
|   | $y_2$ | 0.97 | **0.93** |

*Table 1 Clustering example*

## Scoring

To determine how suspicious an offer in a cluster is, we apply a scoring function to calculate a confidence score for each offer. This scoring function can consider several traits or indicators of "typical" counterfeits. One of the most significant traits of an imitation is the considerably lower price compared to the original product. Although there may be counterfeits having the same price as the original product, based on our experience it is the most important trait to find likely counterfeit offers on the web. This is also because the most important criterion for a consumer to buy fake products is a low price (Schuchert-Güler and Eisend 2003). Further indicators include according to (Schäfer, et al. 2008) and (Jordan and Kutter 2012):

- Retailer/seller rating including user feedback
- Dubious method of payments (e.g., WesternUnion, where a refund is not possible)
- Country of origin of the product (since many counterfeits come from a few countries including China)
- Missing seller or product information
- Missing certifications

- Unusual package sizes (e.g., in medication, where bulk packages suggest a counterfeit, because they are usually not sold to consumers)

- Grammatical and orthographical mistakes in product description

- Type and volume of other products offered by a seller.

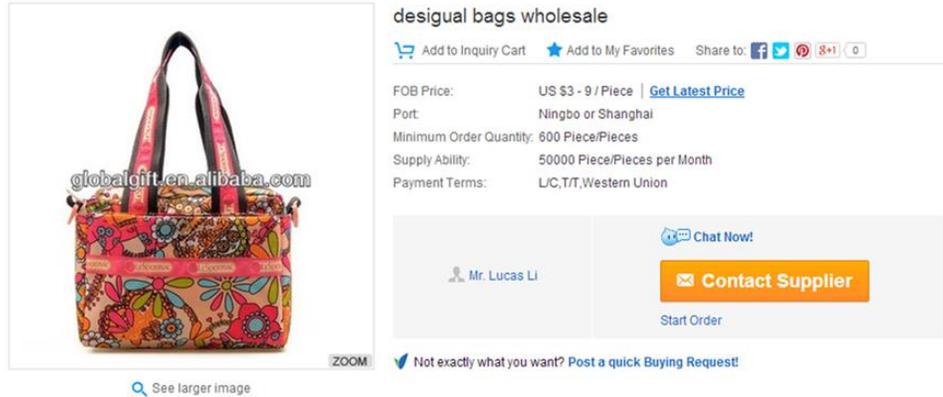

*Figure 2: Suspicious product offer from the platform alibaba.com*

*Figure 2* show an example of a suspicious product offer with some of the mentioned indicators, such as unusual low price, payment method and volume of available products.

By contrast, there exist traits that indicate a reliable offer, e.g., a retailer who has a large product range, a generally good reputation or whose account already exists for a long time. Our approach uses a scoring function which regards some of the criteria listed above and calculates the reliability (trustworthiness) of an offer.

The scoring function is a linear combination of scores based on different indicators with each having a certain weight:

$$s = \frac{w_1*s_1 + w_2*s_2 + \cdots + w_n*s_n}{n} \quad (\sum_i w = 1, \ s_i = [0,1])$$

An indicator for the price can be a function like the following:

$$s_p = \frac{price}{averageOfferPrice}$$

This leads to a score that decreases with price deviations below the average price in a cluster of offers.

For the seller rating we often have a direct value from the data source which can be normalized and used directly. For instance, eBay provides already a rating of sellers in percent. Other sites may use stars which can be mapped to numerical values. Indicators like the payment method or origin can be mapped to numerical values just by categorizing origin countries or payment methods in groups and assigning each group a value between 0 and 1. In the simplest case indicators for counterfeits can just be assigned a one for existing in an offer there and a zero otherwise. The scoring function generally depends on the domain and sales platform.

## Case Study Setup

### Scope

To test the applicability of the described workflow we evaluated it regarding the two steps that influence the quality the most: The clustering step and the scoring step that allows the final categorization in good offers and likely fakes. Without a correct clustering of similar offers we cannot effectively use a scoring function that relies on comparisons between offers of the same type (e.g. one that puts heavy weight on price differences and not just absolute prices). If a satisfactory clustering can be achieved the scoring function has to be evaluated the give insight into the overall quality of the approach. Evaluation of query generation and data extraction is beyond the scope of this initial study. For our evaluation we thus focus on relatively easy to find product offers. We also consider only offers from the eBay platform were we can use a comprehensive API[3] to quickly gather all the information needed.

---

[3] eBay Developers Program, go.developer.ebay.com

For our case study we cannot provide definite answers on whether presumed counterfeit offers really refer to faked products as we do not have the real products and their assessment by experts. For evaluation purposes, we try to manually estimate the correctness of the final scoring step based on the content of the comments belonging to a seller (external and on the site), the overall offer description as well as factors like if traders have been banned at some time after the data was extracted.

**Test data**

In order to judge the effectiveness of our approach, we generated a sample query to see how many suspicious offers were detected. We first defined a specific company and a general product. For the evaluation we focused on several brand products that are often targeted by counterfeiters and offered on the web. We looked at multiple types of clothing products (brand shoes and shirts) and well known perfumes that fall into higher price segments. Overall we extracted 1341 offers (see *Table 2*). Offers refer to a product that can be easily found by querying the product name and manufacturer while applying a certain category as a filter for the output. We decided to use the product type *Eau de Toilette* for the evaluation of the final scoring of offers in regard to the likelihood of being offers for counterfeit goods. These offers and their respective venders were manually flagged as counterfeits or non-counterfeits based on the user comments on the platform and external sites as well as the deletion of offers by the platform owner.

| *Product* | *# of offers* | *# of manually validated counterfeits* |
| --- | --- | --- |
| *Brand Sport Shoes* | *321* | *-* |
| *Brand Shirt* | *233* | *-* |
| *Eau De Toilette* | *787* | *78* |

*Table 2 Test Products and # of offers*

## Clustering and scoring method

For clustering we used the described approach on hierarchical clustering. We use a combination of lexicographic matchers to calculate the score for each element pair between clusters. For our tests we used a minimal pairwise similarity of 0.7.

For the scoring the trustworthiness of offers, we started with a scoring function that regards the price, the retailer rating and the country of origin. Our formula originally looked as follows, with $w_i$ being the weights and $s_i$ being the scores of the several parameters we regard:

$$S = \frac{w_1 * scorePrice + w_2 * scoreRating + w_3 * scoreOrigin}{3} \qquad (\sum_i w = 1, \; s_i = [0,1])$$

For price scoring we use

$$scorePrice = \frac{price}{maxPrice - (maxPrice * 0.25)}$$

Thus, the confidence of an offer decreases linearly with the deviation from the maximum price of the cluster. We declare a small scope below the maximum price as confidential, e.g. 25 %, but products that are cheaper are gradually assigned lower scores. The scoring functions for the retailer is the percentage of positive ratings retrieved for a seller and the function for the origin assigns a score of 0.5 to the top countries frequently being involved in counterfeits and a score of 1 for other countries.

We planned on specifically weighting and testing multiple combinations. However, experiments quickly showed that rating and country of origin cannot be used as reliable scoring factors for the eBay platform using the chosen products. There was no correlation between rating and likelihood of counterfeit offers. Although the seller score provided by eBay appears to be a sensible criterion to judge the confidence of a product offer, it revealed to be rather misleading. First of all, we found possible counterfeit sellers having a perfect score of 100 %, while we also found sellers having a lower score even though their accounts belong to well-known retailers. Note that 98 % is already a very low score on eBay, because most users give positive or neutral feedback.

Taking a closer look on the negative user comments, we found out that most users complained about long delivery times (or even no delivery of the ordered product at all), broken or damaged products, receiving the wrong product (or a wrong size, color etc. of it), impolite or even abusive retailers, troubles when trying to return the product and to obtain the full refund etc. These are apparently the everyday problems users of eBay have to deal with, while only very few complaints are about (possible) fake products. Although it might be assumed that an unreliable retailer might also tend to be a counterfeiter, such a conclusion is quite speculative. Additionally, we often found negative feedback referring to a disappointing product which is however beyond the responsibility of the retailer.

The scoring function generally depends on the domain and sales platform. The importance of indicators for counterfeits on the eBay platform for shoes might be different from indicators for all web shops and a different product. However, the price is an attribute that always exists and at the same time the most important indicator. Because of that we finally limited our tests to essentially using the *scorePrice* function alone.

# Results

## Clustering

*Table 3* shows the effectiveness of the described clustering approach for the 3 different product types. Accuracy in this case refers to the percentage of clusters containing only offers for the same product (all T-shirts of type A in one cluster, T-shirts of type B in a different cluster etc.). Offers for the same product that fall into different cluster are not inherently bad for our approach if the clusters are still large enough. We also specify the number of "superfluous clusters" that are not correct but refer to a single product so that they could be merged with another cluster. On average we obtained about 80 percent correct clusters and about 20 percent of clusters that are superfluous or contain different offers which can lead to wrong scoring of offers.

| Product | # of clusters | # of correct clusters | # of superfluous clusters | Accuracy |
|---|---|---|---|---|
| *Sport Shoes* | 37 | 30 | 4 | 0,81 |
| *Brand Shirt* | 34 | 29 | 2 | 0,85 |
| *EDT* | 54 | 41 | 5 | 0,76 |
| *Overall* | 125 | 100 | 11 | 0,80 |

*Table 3 Clustering quality*

Considering, the high heterogeneity of product offers the achieved clustering thus can be seen a satisfactory. We will also evaluate the quality impact for a corrected clustering where we merge the offers of superfluous clusters with their real clusters.

## Overall quality

To gain insight in the overall quality of the scoring step and the influence of the clustering step we test the counterfeit scoring with both the results of the fully automatic clustering and with a manually corrected clustering.

*Table 4* shows the result of our approach with the fully automatic clustering. The scoring function returns a score between 0 and 1 that denotes how likely it is that an offer is for a counterfeit product. We chose the thresholds 0.6, 0.7, 0.8 and 0.9 for classifying offers into reliable offers if they score above the threshold and counterfeit offers if they score below. *Table 4* shows the precision, recall and f-measure values for each of the threshold values.

The precision is the percentage of offers that were correctly classified as suspicious by our approach:

$$precision = \frac{|\text{offers automatically classified as counterfeits} \cap \text{offers manually classified as counterfeits}|}{|\text{offers automatically classified as counterfeits}|}.$$

By contrast, recall is the ratio of the manually flagged suspicious offers that could be found automatically:

$$recall = \frac{|\text{offers automatically classified as counterfeits} \cap \text{offers manually classified as counterfeits}|}{|\text{offers manually classified as counterfeits}|}.$$

The F-Measure is the harmonic mean of precision and recall.

| Threshold | Recall | Precision | F-Measure |
|---|---|---|---|
| 0,6 | 0,064 | 0,833 | 0,119 |
| 0,7 | 0,564 | 0,543 | 0,553 |
| 0,8 | 0,564 | 0,283 | 0,378 |
| 0,9 | 0,564 | 0,22 | 0,317 |

*Table 4 Results after automatic clustering step*

It can be seen that a high (reliability) threshold leads to more offers falling below it and being classified as not reliable. The result then includes a high amount of the manually flagged offers thereby improving recall. By contrast, precision improves for lower thresholds as these reduce the likelihood of false counterfeit candidates. The threshold of 0.7 produces the best balance resulting in the highest F-Measure value. At this point the recall stays stable compared with higher thresholds while the precision is already at a good level.

*Table 5* shows the results when the clustering is manually corrected. In this case, precision is perfect for the 0.6 threshold, i.e., there are no false positives in the offers classified as counterfeits. This is made possible by the correct clustering so that only the prices for offers of the same product are compared with each other and threshold 0.6 refers to unrealistically large price differences. The best F-Measure is again achieved for threshold 0.7. Compared to the automatic clustering case (*Table 4*) we observe the same recall but an improved precision due to the corrected clustering. The overall F-Measure improved from 55 to about 60%.

| Threshold | Recall | Precision | F-Measure |
|-----------|--------|-----------|-----------|
| 0,6 | 0,051 | 1 | 0,098 |
| 0,7 | 0,564 | 0,628 | 0,595 |
| 0,8 | 0,564 | 0,4 | 0,468 |
| 0,9 | 0,564 | 0,244 | 0,341 |

*Table 5 Results for manually corrected clustering*

Altogether, 10.3 % of the offers (81 of 787) were flagged as suspicious fully automatically when using threshold 0.7. In our case study we found out that 54.3 % of these offers (44) are most likely real counterfeit offers. While the evaluation results are not yet perfect, they show the viability of the proposed approach as it is possible to let the user only verify a smaller subset of the offers (81 instead of 781 offers) which can mean a significant time savings for counterfeit detection.

One interesting result was that suspicious retailers who appeared in more than one cluster, had offers with very bad scores in every cluster. For instance, there was one highly suspicious retailer having offers in 9 clusters. The scores of the offers were very low in every cluster. This indicates it may be a good idea to use such collected evidence about suspicious retailers in future scorings.

## Conclusion & Future Work

We proposed a new approach to semi-automatically identify offers of counterfeits in online sales platforms such as large auction sites or web shops. The approach is based on a workflow with automatic generation of search queries for the products of interest as well as a clustering and scoring the trustworthiness of product offers. The initial evaluation showed the viability of the proposed approach but also the need for further improvements. The proposed clustering scheme worked relatively well but can be manually corrected by merging several clusters with offers for the same product. The manual verification of likely counterfeits can be restricted to smaller subsets of the product offers thereby limiting the effort for counterfeit identification.

In future work, we see the need to evaluate and fine-tune the propose approach for additional sales platforms and product types. The used scoring based on price is only a first step and should be extended with additional criteria based on product type, sales platform and perhaps insights from initial evaluations, e.g. about suspicious retailers.

An inherent limitation remains the difficult manual decision about whether a suspicious offer is really about a faked product. Hence the manual verification and fine-tuning should ideally be performed in collaboration with experts affected by counterfeits such as the manufacturers of frequently faked products or platform owners.